\documentclass[aps,amssymb,amsfonts,floatfix,showpacs]{revtex4}

\usepackage{graphicx}
\usepackage{bm}
\usepackage{mathbbol}

\begin{document}

\title{Transport and Control in One-Dimensional Systems}

\author{Lea F. Santos} \thanks{Email: {\tt lsantos2@yu.edu}}
\affiliation{\mbox{Department of Physics, Yeshiva University, 245 Lexington Ave, New York, NY, 10016, USA}}

\date{\today}

\begin{abstract}
We study transport of local magnetization 
in a Heisenberg spin-1/2 chain at zero 
temperature. The system is initially prepared in 
a highly excited pure state far from equilibrium and its evolution is
analyzed via exact diagonalization. 
Integrable and non-integrable regimes
are obtained by adjusting the parameters of the Hamiltonian, 
which allows for the comparison of transport behaviors in both limits.
In the presence of nearest neighbor
interactions only, the transport behavior in the integrable clean system contrasts with 
the chaotic chain with on-site defects, oscillations
in the first suggesting ballistic transport and a fast decay in the latter indicating
diffusive transport. The results for a non-integrable system with frustration
are less conclusive, similarities with the integrable chain being verified.
We also show how methods of quantum control may be applied to
chaotic systems to induce a desired
transport behavior, such as that of an integrable system.
\end{abstract}

\date{\today}
\maketitle

\section{Introduction}

The understanding of the behavior of quantum many-body systems far from equilibrium 
is still far from satisfactory. Under debate, for example, are the conditions 
that lead to the thermalization of isolated quantum 
systems~\cite{Manmana2007,Rigol2007,Rigol2008} or
to ballistic and diffusive transport.
The distinction between ballistic and 
diffusive transport in linear response theory 
usually relies on the computation of the Drude weight, which 
is finite in the first case~\cite{Zotos1999,Zotos2005},
but alternatives approaches have also been considered.
In studies of thermal transport, for instance, the derivation of the Fourier law from a
microscopic foundation has been pursued by applying the
Hilbert space average method~\cite{Michel} or by numerically simulating 
open systems coupled to heat reservoirs~\cite{Saito,Mejia}. 

In addition to fundamental questions, the subject of
quantum transport received increased attention 
after experimental observations of unusual excess conductivity for 
heat~\cite{Kudo1999,Sologubenko,Hess,Hess2007}
and magnetization~\cite{Takigawa1996}
in low-dimensional magnetic compounds. 
Advances in optical lattices~\cite{Greiner2008}
are also very promising, since these highly controllable systems 
allow for the investigation of the dynamics 
of strongly correlated quantum systems for long times.
Ultracold atom gases have been realized
in optical lattices~\cite{Paredes2004} and  magnetic systems
may soon be implemented~\cite{Trotzky2008,BarthelARXIV}.
Magnetic compounds are well described by Heisenberg spin-1/2 systems, which 
in part explains the proliferation of articles analyzing transport in these systems.
Under certain conditions, 
this model may also be mapped into spinless fermion systems via 
a Jordan-Wigner transformation~\cite{Jordan1928} or onto systems of 
hard-core bosons~\cite{Rigol2007}.

A special feature of the one-dimensional Heisenberg spin-1/2 model in the absence of defects and with nearest-neighbor interactions only is 
that the model is integrable and solved with the Bethe Ansatz method~\cite{Bethe}.
The addition of impurities~\cite{Avishai2002,Santos2004} or
next-nearest neighbor interactions~\cite{Hsu1993,Kudo2005} 
to the chain may lead to the onset of quantum chaos.
By varying the parameters of the Hamiltonian we may then analyze transport behavior 
in both regimes.
At finite temperature, in the gapless phase, one generally associates integrability with ballistic transport and chaoticity with 
diffusion~\cite{Zotos1997,Alvarez2002,Heidrich2003,Heidrich2004,Rabson2004,Zotos2005,Steinigeweg2006,Mukerjee2008}. 
In the gapped phase, diffusive behavior is found also in integrable systems~\cite{Shastry1990,Narozhny1998}.
However, there are still open questions, especially about the transport in integrable systems at finite~\cite{Narozhny1998,Mukerjee2008} and
infinite temperatures~\cite{Fabricius1998}, and 
in nonintegrable systems at low
temperatures~\cite{Rosch2000,Jung2006,LangerARXIV} - such as the possibility 
of ballistic transport in frustrated 
chains~\cite{Narozhny1998,Kirchner1999,Alvarez200289,Fujimoto2003,Heidrich2004,Zotos2004,Heidarian2007,Mukerjee2008}. At zero temperature, within linear
response theory, the correspondence `gapless phase -- ballistic
transport' and `gapped phase -- diffusive behavior' 
is well established~\cite{Shastry1990,Bonca1994}, but a new
discussion has emerged concerning the problem
of quench dynamics in the case of initial states very far from equilibrium.
The latter is the focus of the present work.

Here, we compare transport of local magnetization in an isolated 
Heisenberg spin-1/2 chain with free boundaries
in both integrable and non-integrable regimes.
A highly excited initial state
corresponding to spins pointing up in the first 
half of the chain and  pointing down in the other half is considered. 
We assume, as in Ref.~\cite{Steinigeweg2006}, that an exponential decay of the magnetization
to equilibrium indicates diffusive transport and that a bouncing behavior indicates ballistic transport.
Two chaotic chains are considered: clean with frustration
in the three directions
and disordered with a single defect in the middle of the chain.  
Exact diagonalization 
is used, which limits the analysis to small system sizes.
An exponential relaxation to equilibrium is verified for the disordered system, 
while the frustrated chain shows oscillations similar to those seen in the 
integrable system~\cite{Santos2008PRE}. It has been suggested that such 
oscillations may not be related to ballistic transport~\cite{LangerARXIV}, 
but are possibly related 
to the quantum phase transition of the isotropic 
chain~\cite{BarthelARXIV,BarmettlerARXIV}. 
We show that even though reduced, the oscillations persist
also for anisotropic frustrated systems in the gapless phase.
We argue that the similarities between frustrated and integrable chain result from the mixing
of different symmetry sectors, 
which does not occur in the dynamics of the disordered system.
In the gapped phase, localization of local magnetization is
observed for all systems, as expected.

In the second part of this work, we discuss how methods of quantum control may be 
used to induce a transition from diffusive to ballistic transport.
This is accomplished by applying a sequence of control operations which eliminate
unwanted terms of the Hamiltonian~\cite{ErnstBook,HaeberlenBook,Viola1998,Viola}.
The system studied here is subjected to a static magnetic field in the $z$ direction;
all sites having the same Zeeman splitting, except the defect.
By applying a sequence of very strong magnetic fields that frequently rotate
all spins perpendicularly to the $z$ direction, we show that the
effects of the on-site disorder may be averaged out and the transport behavior of the integrable chain
recovered.
A sequence that eliminates the effects of next-nearest neighbor interactions has also 
been proposed in~\cite{Santos2008PRE}.

The paper is organized as follows. Sec.~II describes the model
and the quantity used to characterize chaos.
Sec.~III shows the results for local magnetization in 
the gapless and gapped phase for both integrable and
non-integrable regimes.
Sec.~IV compares the transport behavior of the disordered chain in the 
absence and presence of a control sequence.
Discussions and concluding remarks are presented in Sec.~V.

\section{System Model and Characterization}

We consider a one-dimensional Heisenberg spin-1/2 system with open boundary conditions described by the
Hamiltonian,

\begin{eqnarray}
H &=& H_z + H_{NN} +  H_{NNN} \nonumber \\
&=& \sum_{n=1}^{L} \omega_n S_n^z  + \sum_{n=1}^{L-1} J 
(S_n^x S_{n+1}^x + S_n^y S_{n+1}^y + \Delta S_n^z S_{n+1}^z) 
+ \sum_{n=1}^{L-2} J' 
(S_n^x S_{n+2}^x + S_n^y S_{n+2}^y + \Delta S_n^z S_{n+2}^z) \:,
\label{ham}
\end{eqnarray}
where $\hbar=1$, $L$ is the number of sites, and $S_n^{(x,y,z)}$ are the spin operators at site $n$
in the three directions.
The parameter $\omega_n$ is the Zeeman splitting 
of spin $n$ as determined by a static magnetic field in the $z$ direction. 
The system is clean when all sites have the same energy splitting $\omega_n=\omega$, 
and it is disordered when defects characterized by different energy splittings 
$\omega_n=\omega+d_n$ are present.
$J$ and $J'$ are the exchange coupling strengths 
of nearest-neighbor (NN)
and next-nearest-neighbor (NNN) couplings, respectively, and are assumed to be constant;
$\Delta $ is the anisotropy associated with the
Ising interaction $S_n^z S_{n+1}^z$. At zero temperature,
$-1<\Delta\leq 1 $ corresponds to the gapless phase and $|\Delta|>1$
to the gapped phase. 
We set $J, J', \Delta>0$.
In the following, a spin up may sometimes be referred to as an excitation.

The total spin operator in the
$z$ direction, $S^z=\sum_{n=1}^L S_n^z$, is conserved, so
the matrix $H$ is composed of independent blocks, each
of dimension $N={L \choose M}=L!/[(L-M)!M!]$, where $M$ is the total number of
spins pointing up. We consider $L$ even and study the largest subspace, $S^z=0$, which
corresponds to $M=L/2$ (from the perspective of 
interacting spinless fermions, we are at half-filling). 
The calculations are performed in the basis consisting 
of eigenvectors of $S^z$. 

In quantum systems, integrable and non-integrable regimes may be identified by analyzing the 
distribution of spacings $sp$ between neighboring energy levels 
\cite{HaakeBook,Guhr1998}. 
Quantum levels of integrable systems tend to cluster and are not prohibited from crossing, the typical distribution is Poissonian: $P_{ P}(sp) = \exp(-sp)$.
In contrast, non-integrable systems show levels that are correlated and crossings are avoided, the level statistics is given by the Wigner-Dyson distribution. The form of the Wigner-Dyson distribution
depends on the symmetry properties of the Hamiltonian. In the case of systems with time reversal invariance we have: $P_{ WD}(sp) = (\pi sp/2)\exp(-\pi sp^2/4)$. The Heisenberg model
with a magnetic field, as in $H$~(\ref{ham}), does not commute with the conventional
time-reversal operator, but the 
Hamiltonian elements are real and the distribution is still given by 
$P_{ WD}(sp)$~\cite{HaakeBook,Brown2008}.

\begin{figure}[htb]
\includegraphics[width=5in]{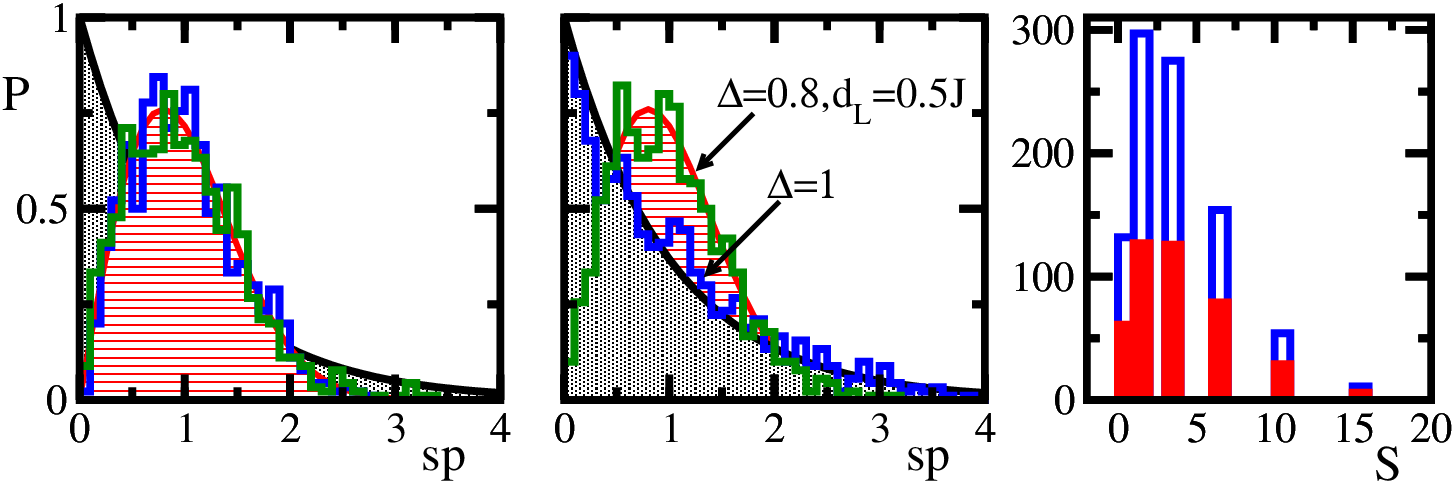}
\caption{(Color online.) Onset of chaos 
in a Heisenberg chain with $L=12$.
Left panel: Disordered chain with NN interactions only, as
described by $H$~(\ref{ham}) with $J'=0$ and $d_{L/2+1}=0.5J$.
Both $\Delta=1$ and $\Delta=0.5$ lead to distributions similar to $P_{WD}$.
Dotted region -- $P_{P}$; Horizontal lines -- $P_{WD}$.
Middle panel: Frustrated chain, as
described by $H$~(\ref{ham}) with $J'=J$. The chaotic system with $\Delta=1$ and $d_n=0$
gives a Poisson distribution when $S^2$ and reflection symmetries are mixed.
A Wigner-Dyson distribution is recovered when both symmetries are broken by 
considering $\Delta=0.8$ and $d_L=0.5J$.
Right panel: Histogram of eigenstates of $H$~(\ref{ham}) with $J'=J$ and $\Delta=1$
according to their $S^2$ subspaces. Empty bars: all eigenstates; filled bars: eigenstates
$j$ that contribute
to $|\Psi(t)\rangle$ with probability amplitude $|a_{j,in}|\geq 10^{-4}$.}
\label{fig:eta}
\end{figure}

A meaningful level spacing distribution requires the separation
of different symmetry sectors. If energies from different subspaces are mixed, we may
obtain a Poisson distribution even if the system is chaotic. The main feature
of chaoticity, level repulsion, is therefore lost.
In addition to $S^z$, $H$~(\ref{ham}) conserves also
parity, when the system is clean, and 
total spin $S^2=(\sum_{n=1}^L \vec{S}_n)^2$, when the system is
clean and isotropic, $\Delta=1$. A particular feature of 
the $S^z=0$ subspace is its invariance also under a $\pi$-rotation of all
spins~\cite{Brown2008}. This symmetry is taken into account when $S^2$ 
is considered, but it remains for an anisotropic system~\cite{Kudo2005}.
In the case of a disordered chain with NN interactions only, 
a single defect placed in the middle of the chain, $d_{L/2+1}\neq0$, may
lead to the onset of chaos~\cite{Santos2004}, as shown
in the left panel of Fig.~\ref{fig:eta}. A clean system with $J=J'$ and $\Delta=1$ is also
chaotic~\cite{Hsu1993}, but it conserves $S^2$ and parity. If these
symmetries are not taken into account, a misleading Poisson distribution
is obtained, as shown in the middle panel of Fig.~\ref{fig:eta}. How these
states distribute among the various $S^2$ subspaces is given in the right panel
for $L=12$; moreover both parities are present, from the
$N=924$ eigenstates, 472 have odd parity. The $S^2$ symmetry may be broken 
by choosing $\Delta \neq 0$, but here spin reversal and translation
invariance remain. The three symmetries are guaranteed to 
be broken by placing a defect
in one of the edges of the chain~\cite{Santos2004,SantosEscobar2004}, 
an alternative that does not affect the integrability of the 
clean system with NN-coupling~\cite{Alcaraz1987}. 
By combining both, defect and anisotropy, 
a clear Wigner-Dyson distribution is reached for the 
system with NNN interactions, as shown in the middle panel.

\section{Transport Behavior: Local Magnetization}

We study transport of local magnetization as defined by

\begin{equation}
M (t) \equiv \langle \Psi(t) | \sum_{n=1}^{L/2} S_{n}^z| \Psi(t) \rangle,
\label{mag}
\end{equation}
where $| \Psi(t) \rangle$ 
is the state of the system at instant $t$.

The initial state $| \Psi(0) \rangle$ considered here is 
the basis vector
$|\phi_{in} \rangle$ of $S^z$ corresponding to 
spins pointing up in the first half of the chain 
and pointing down in the other half.
The eigenvectors $| \psi \rangle$ of $H$~(\ref{ham}) in the basis $|\phi \rangle$ are
given by $| \psi_j \rangle = \sum_{k} a_{j,k} |\phi_k \rangle$
with eigenvalues $E_j$, so
the local magnetization may be written as

\begin{equation}
M (t) = \frac{1}{2} \sum_{j=1}^N \left[ 
\left( \sum_{n=1}^{L/2} \xi_{j,n} \right)
\left( 
\sum_{k=1}^N (a_{k,in} a_{k,j})^2 +
2 \sum_{p=1}^{N-1} \sum_{q=p+1}^{N} 
a_{p,in} a_{p,j} a_{q,in} a_{q,j} \cos [(E_p - E_q)t]
\right)
\right],
\label{cos}
\end{equation}
where $\xi_{j,n}=-1$ (+1) if spin $n$ of state $|\phi_j \rangle$ is pointing down (up).

The scenario is that of a quench dynamics, where the system is initially prepared
in the eigenstate $| \Psi(0) \rangle=| \uparrow \uparrow \ldots \uparrow \downarrow \ldots 
\downarrow \downarrow \rangle$ of $H_z$ and then suddenly interactions are turned on and the system
dynamics becomes dictated by $H$~(\ref{ham}).

\begin{figure}[htb]
\includegraphics[width=4.0in]{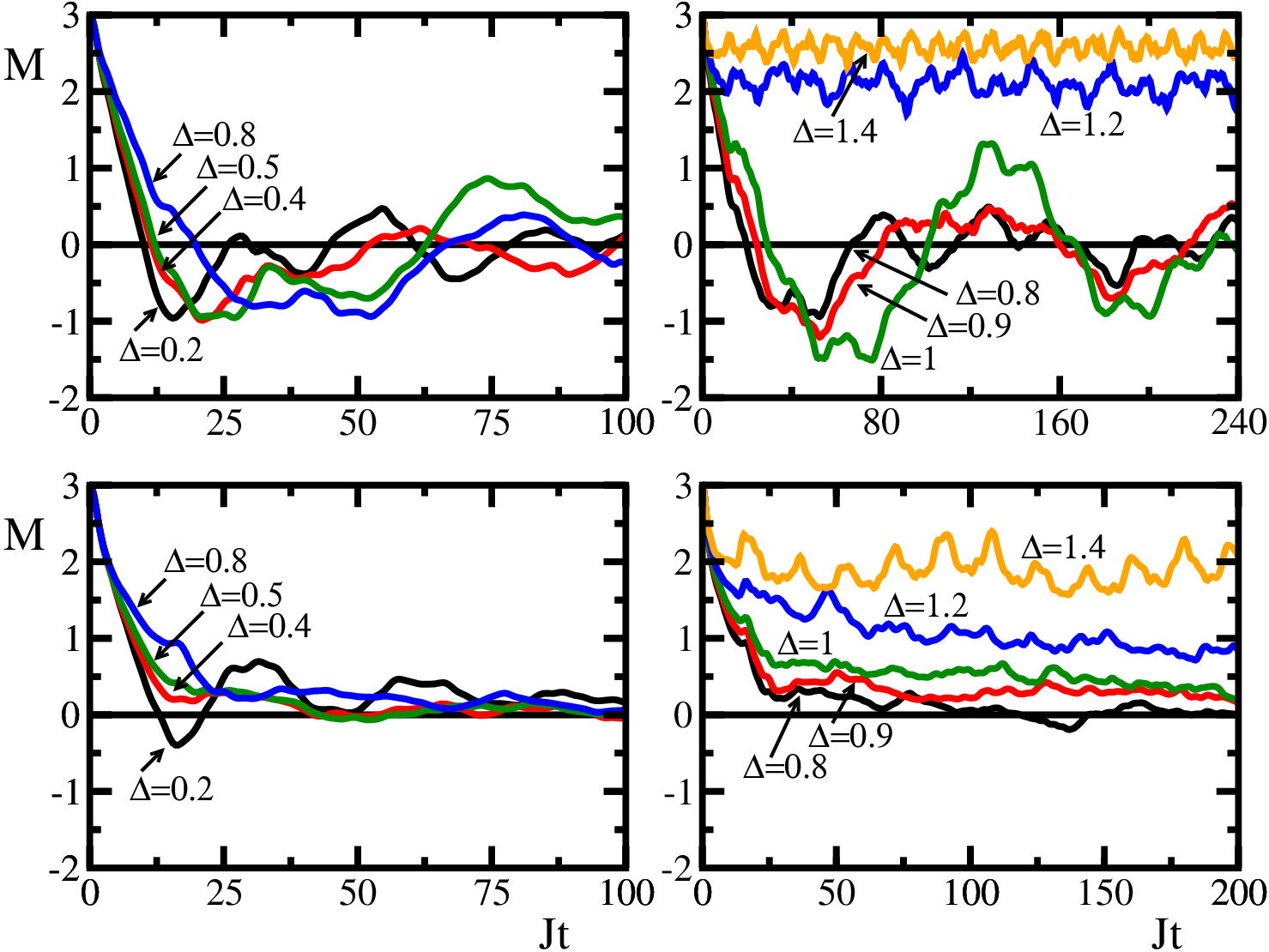}
\caption{(Color online.) Transport of 
local magnetization in a Heisenberg chain with NN interactions 
described by $H$ (\ref{ham}) with $J'=0$ and $L=12$.
The value of $\omega$ is irrelevant for the
dynamics and $\Delta$ is indicated. 
Top panels: $M(t)$ for the clean integrable system with $d_n=0$.
Bottom panels: $M(t)$ for the chaotic disordered system with $d_{L/2+1}=0.5J$.}
\label{fig1}
\end{figure}

{\em Integrable vs. disordered chain.} The 
clean system with NN interactions only, 
as described by $H$~(\ref{ham}) with $d_n=0$ and $J'=0$, is
an integrable model. 
The dynamics of local magnetization for
different values of $\Delta$ is shown in the top panels of Fig.~\ref{fig1}. 
When $\Delta>1$, few states are in resonance
with $|\phi_{in} \rangle$ and therefore
localization occurs (right panel). In the gapless phase, the bouncing behavior 
of $M(t)$ suggests ballistic 
transport~\cite{Steinigeweg2006}. However, it calls attention 
the reduction of the oscillations amplitudes as
$\Delta$ decreases (left panel), which is probably associated with the breaking of 
the $S^2$ symmetry.

The addition of on-site disorder to the 
chain with NN interactions breaks $S^2$ symmetry and parity.
A single defect $d_{L/2+1}=0.5J$  
leads to the onset of chaos~\cite{notePRE} and to a fast
decay of the local magnetization to equilibrium for $0.4\lesssim \Delta\lesssim 0.8$,
as seen in the bottom left panel of Fig~\ref{fig1}.
As $\Delta $ increases -- bottom right panel -- defect and anisotropy 
reduce resonances with the initial state 
and localization eventually takes place. 

\begin{figure}[htb]
\includegraphics[width=4.0in]{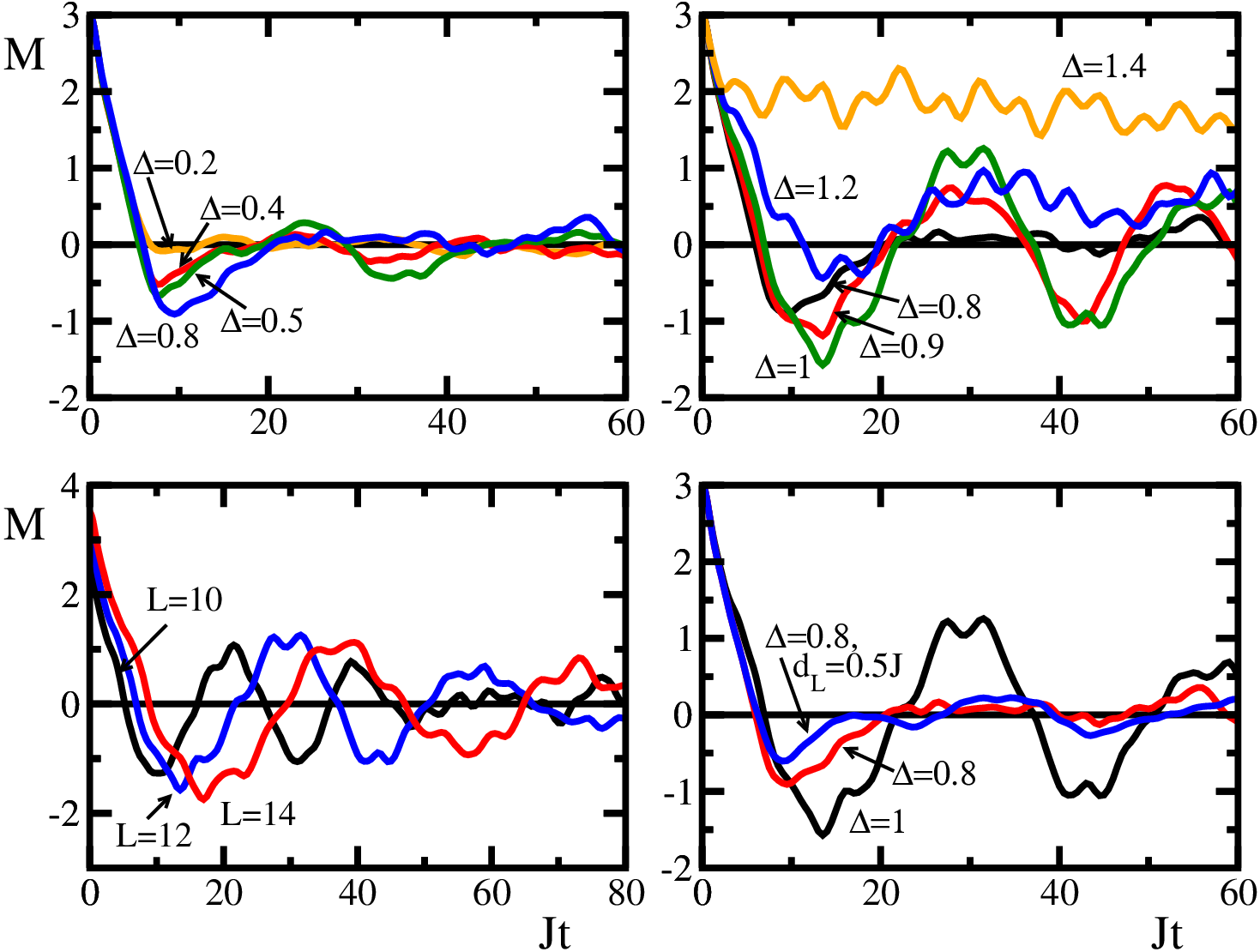}
\caption{(Color online.) Transport of 
local magnetization in a Heisenberg chain with NNN interactions
described by $H$~(\ref{ham}) with $J'=J$ and $L=12$. 
Top panels: $M(t)$ for the frustrated system with different values of $\Delta$.
Left bottom panel: Comparison of  
$M(t)$ for different system sizes, $\Delta=1$. Right bottom panel: Approach to diffusive
transport as symmetries are broken. Present symmetries: 
$S^2$ and parity when
$\Delta=1$ ; parity when $\Delta=0.8$; none of the two when
$\Delta=0.8$ and $d_L=0.5J$.}
\label{fig2}
\end{figure}

{\em Integrable vs. frustrated chain.}
$H$~(\ref{ham}) with $d_n=0$ and
$J=J'$ describes a chaotic system.
Instead of diffusive transport, however, 
the plots in Fig.~\ref{fig2} indicate the following.
(i) Localization occurs again, as expected, when $\Delta>1$.
(ii) When $\Delta \sim 1$, the evolution of local magnetization shows a bouncing behavior  
that is similar to that observed for the integrable system.
The oscillations for $J=J'$ are faster than for $J'=0$,
but their amplitudes are of the same order (cf. top right panels of Figs.~\ref{fig1} and \ref{fig2}).
The bottom left panel 
suggests that these oscillations are not artifacts of the small size of the chain, since
their amplitudes do not appear to decrease as the system size increases, although 
simulations for larger systems are necessary to settle this 
point~\cite{noteHastings}.
(iii) According to the top left panel of Fig.~\ref{fig2}, the 
reduction of oscillations for the system with NNN interactions associated with the
decrease of anisotropy, $\Delta <0.8$, is more significant than in the 
integrable case (cf. top left panels of Figs.~\ref{fig1} and \ref{fig2}).
Notice that the curve for the frustrated chain with $\Delta=0.2$ is hardly noticeable due to
the absence of oscillations and the fast decay of $M(t)$ to 
equilibrium.

We interpret observations (ii) and (iii) as follows. 
State $|\Psi(t)\rangle$ is a linear combination of eigenstates of $H$ belonging to a single
$S^z$ sector, but no restriction exists with respect to $S^2$, 
global $\pi$ rotation, and parity.
Similarly to what is observed for 
the level spacing distribution (middle panel of Fig.~\ref{fig:eta}),
but now from a dynamical point of view, the effects of level 
repulsion may be missed by the local
magnetization if those symmetries are present. 
The mixing of symmetry sectors implies that a large fraction of (almost) degenerate states 
contribute to $M(t)$, such level crossings
must reduce the phase randomization that leads to a fast decay of 
the local magnetization~(\ref{cos}). 
For instance, when $L=12$ and $J'=J$, there are 429 eigenstates 
$|\psi_j\rangle $ participating in $|\Psi(t)\rangle$ with probability amplitude
$|a_{j,in}|\geq 10^{-4}$. 
From 400 states (the border ones being neglected), there are
29 unfolded spacings between neighboring levels with $sp \leq 0.1$.
These 429 eigenstates are distributed among the $S^2$ subspaces with quantum  
number $S$ between 0 and 21 as shown in the right panel of Fig.~\ref{fig:eta}, and 
both parities are found, 217 states having odd parity.  
This is to be contrasted with the disordered system, where $S^z$ is the only conserved
quantity. $M(t)$ captures well the transition to chaos and diffusive transport becomes evident.
For $L=12$, $J'=0$, $\Delta=1$, and $d_{7}=0.5J$, there are 751 eigenstates
$|\psi_j\rangle $ participating in $|\Psi(t)\rangle$ with probability amplitude
$|a_{j,in}|\geq 10^{-4}$. From the 720 states in the bulk,
only 5 lead to $sp \leq 0.1$, a percentage 10 times smaller than in the frustrated chain.

For the frustrated chain, as $\Delta$ decreases from 1 and 
the $S^2$ symmetry is broken, 
transport of local magnetization becomes closer to a diffusive behavior.
However, since parity and invariance under a
global $\pi$ rotation are still present,
the decay of $M$ is not as
abrupt as in the disordered system. Better 
agreement is verified if these symmetries are also broken by adding, for example, a
defect to one of the borders of the chain, as
in the bottom right panel of Fig.~\ref{fig2}. This behavior should 
again be compared with the middle panel of Fig.~\ref{fig:eta}, where
$P(sp)$ approaches a Wigner-Dyson distribution when the
symmetries are broken.

\section{Manipulation of Transport Behavior via Quantum Control}

We have recently discussed the use of quantum control methods,
such as dynamical decoupling (DD), to induce a desired
transport behavior~\cite{Santos2008PRE}. DD techniques consist of 
sequences of external control operations. The goal is to average out the 
effects of unwanted terms and reshape the Hamiltonian in order to achieve 
a desired dynamics. 
These methods were first applied in Nuclear Magnetic Resonance (NMR) spectroscopy, where
the control operations correspond to very strong magnetic fields (pulses) able to rotate the spins and time-reverse the system evolution~\cite{HaeberlenBook,ErnstBook}.
More recently their applications have been extended to more general
systems, especially in the context of quantum information, where the purpose is not only
control of internal interactions~\cite{Viola1999Control}, 
but also of external couplings~\cite{Viola1998,Viola1999Dec}. 

Pulse sequences have been developed and used for several different
systems. Sequences to eliminate 
NNN interactions or on-site disorder from $H$ (\ref{ham}) were proposed in Ref.~\cite{Santos2008PRE}.
Consider, for example, the chaotic system with on-site disorder, be it a single defect or various
impurities spread through the chain~\cite{Santos2008PRE}. 
In order to eliminate the effects of the 
disorder and recover the transport behavior of the integrable system, we apply sequences
of instantaneous pulses given by

\[
P_x = \exp \left( -i \pi \sum_{n=1}^{L} S_n^x \right) = \exp(-i \pi S^x),
\]
which rotate all spins 180$^o$ around the $x$ direction.
The pulses at instants $t_{j-1}$ and $t_j$ 
are separated by a time interval $\tau = t_j - t_{j-1}$ of free evolution. They 
alternate the sign of the one-body terms, but 
do not change the two-body interaction terms.
The propagator at time
$t_{2p}=p T_c$, where $p \in \mathbb{N}$
and $T_c=2\tau$ is the cycle time, approaches that of an integrable clean
system as $\tau \rightarrow 0$

\[
U(pT_c)\stackrel{\tau \rightarrow 0}{\rightarrow} \exp[-i H_{NN} p T_c].
\]
This result is achieved as follows. Consider the notation

\[
U_{+} =\exp [-i (H_z + H_{NN}) \tau] = \exp[-i H_1 \tau],
\]
\begin{eqnarray}
U_{-}&=&P_x U(t_j,t_{j-1}) P_x = P_x (P_x P_x^{\dagger}) U(t_j,t_{j-1}) P_x \nonumber \\
&=& -\mathbb{1} \exp \left[ -i \left( e^{+i \pi S^x} (H_z + H_{NN}) e^{-i \pi S^x} 
\right) \tau \right] \nonumber \\
&=& -\exp [-i ( - H_z + H_{NN}) \tau] = -\exp[-i H_2 \tau],
\nonumber
\end{eqnarray}
where $\mathbb{1}$ is the identity operator. The propagator at $t_{2p}$ can then be written 
as

\begin{eqnarray}
U(pT_c) &=& P_x U(t_{2p},t_{2p-1}) P_x  \ldots  P_x U(t_2,t_1) P_x U(t_1,0)\nonumber \\
&=& U_{-} U_{+} \ldots U_{-} U_{+} = \exp[-i \bar{H} p T_c].
\label{DD_sequence}
\end{eqnarray}
The average Hamiltonian, $\bar{H} =\sum_{k=0}^{\infty}{\bar
H}^{(k)}$, is obtained by using the Baker-Campbell-Hausdorff 
expansion~\cite{HaeberlenBook,ErnstBook}.
The first three dominant terms are

\begin{eqnarray*}
&& {\bar H}^{(0)}  = \frac{\tau}{T_c} 
(H_1 + H_2) = H_{NN} ,
\\
&& {\bar H}^{(1)}  = -\frac{i\tau^2}{2T_c} 
[H_2,H_1] ,
 \\
&& {\bar H}^{(2)}  = -\frac{\tau^3}{12T_c}  
\Big\{ [H_2,[H_2,H_1] + [H_2,H_1],H_1] \Big\} .
\end{eqnarray*}

\begin{figure}[htb]
\includegraphics[width=3.5in]{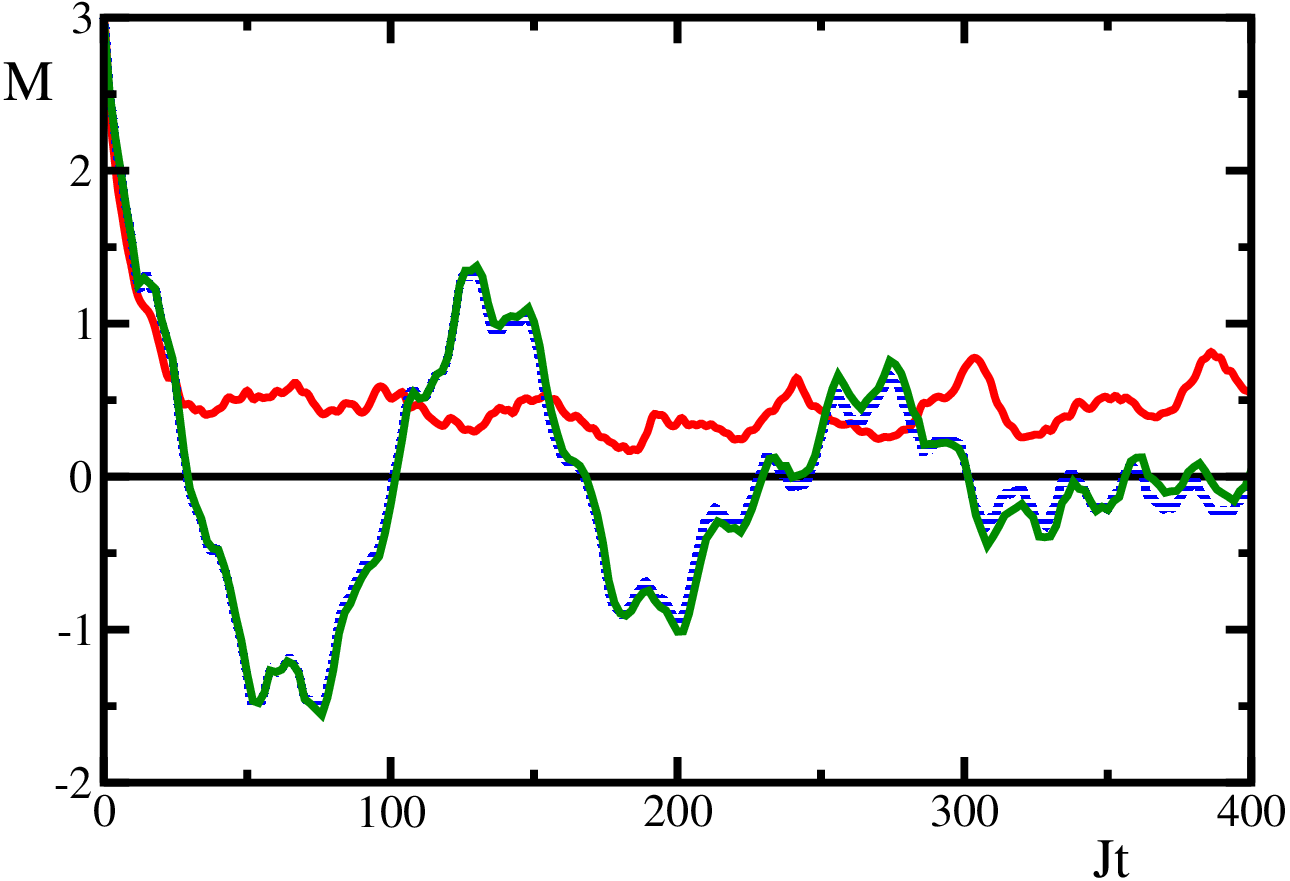}
\caption{(Color online.) Transport of 
local magnetization in a Heisenberg chain 
described by $H$ (\ref{ham}) with $J'=0$, $\Delta=1$, and $L=12$.
The curve showing a fast decay of $M(t)$ corresponds to a disordered system
with $d_7=0.35J$ in the absence of pulses.
Bouncing curves represent the clean chain (dashed line) 
and the disordered system (full line) subjected to the DD sequence (\ref{DD_sequence}).
Data is acquired after every $T_c=2\tau$, where $\tau = J^{-1}$.}
\label{fig3}
\end{figure}

As $\tau \rightarrow 0$,
the average Hamiltonian then approaches 
$H_{NN}$ and the transport of local magnetization 
coincides with the one obtained for the integrable chain.
For small values of $\tau$, the agreement may hold 
for relatively long times, as shown in
Fig.~\ref{fig3}. In real scenarios, however, we need to take into account 
several pulse imperfections and the limitation on pulse separations, which
leads to the accumulation of residual averaging errors. Different 
approaches exist to handle these problems (see Ref.~\cite{Santos2008}
and references therein).

\section{Conclusion}

We compared the transport behavior of local magnetization in 
integrable and chaotic one-dimensional many-body systems 
in the case of an initial state far from equilibrium and studied 
quantum control methods as possible tools to manipulate transport behavior.
An isolated Heisenberg spin-1/2 chain 
was the study-case. Chaos was induced by adding 
on-site disorder or frustration to the system.

We found that in the gapped phase, local magnetization is localized for all systems. In the
gapless phase, diffusive transport is evident in the disordered chain, but the frustrated
chain shows a behavior similar to the integrable model, especially when closely  
isotropic. We argued that this is a consequence of the mixing of different symmetry
sectors, which occurs in the dynamics of 
both the frustrated and the integrable system. The combination
of different subspaces obfuscates the effects of level crossing, which is a main
feature of chaotic systems, and slows down
the decay of local magnetization to equilibrium.
These results indicate that a one to one correspondence between quantum chaos and diffusive
transport does not necessarily hold. To be able to view transport as a 
dynamical signature of chaos, we need to consider quantities that reside in a single
symmetry sector. Local magnetization, for instance, is such a quantity when parity
and total spin are not conserved, as in the disordered system, 
since it resides in a single subspace of total spin in the $z$ direction.

We also discussed a sequence of control operations that may be applied to a disordered
system in order to remove the effects of the defects and recover ballistic transport. The possibility
to manipulate transport behavior via quantum control methods may find practical applications,
such as the design of devices with controllable conductivity, the mitigation of the effects of local heating, or the transport of information in quantum computers~\cite{Cappellaro2007,Cappellaro2007b}.
Real systems where these ideas might be tested include crystals of 
fluorapatite~\cite{Cappellaro2007,Cappellaro2007b}, as 
studied in NMR, or optical lattices~\cite{Paredes2004,Greiner2008,Trotzky2008}.

\begin{acknowledgments}
We thank C. O. Escobar for useful discussions. 
This research was supported by an award from Research Corporation.
\end{acknowledgments}

\end{document}